\documentclass{elsart}

\usepackage{natbib}

\usepackage{amssymb}

\newcommand{\be}{\begin{equation}}
\newcommand{\ee}{\end{equation}}
\newcommand{\bea}{\begin{eqnarray}}
\newcommand{\eea}{\end{eqnarray}}

\begin{document}

\begin{frontmatter}


 \title{Inflationary parameters and primordial perturbation spectra\thanksref{label1}}
 \thanks[label1]{Invited talk at CMBNET meeting, Oxford, February 2003}
 \author{David Wands}
 \ead{david.wands@port.ac.uk}
\address{Institute of Cosmology and Gravitation, University of
  Portsmouth,\\Portsmouth, PO1 2EG, United Kingdom}




\begin{abstract}
  I discuss how parameters describing inflation in the very early
  universe may be related to primordial perturbation spectra.
  Precision observations of anisotropies in the cosmic microwave
  background (CMB) such as those provided by the WMAP satellite offer an
  unprecedented window onto the physics of the very early universe.
  To theorists exploring speculative models of physics at high
  energies, the CMB offers us the chance to put our ideas to the test.
\end{abstract}


\end{frontmatter}


\section{Introduction}
\label{intro}

Inflation was originally proposed to solve classical problems of the
hot big bang ``standard model''. Inflation sought to explain the
remarkable homogeneity on large scales, negligible spatial flatness,
the large entropy of the observable universe and the absence of
monopoles and other dangerous relics from phase transitions in the
young universe. But it was the realisation that it also provides a
quantum origin for the large scale structure of the universe
that has proved to be its greatest success. This
was an unexpected bonus that puts models of inflation in the very
early universe into the realm of testable science and not simply a
matter of philosophical prejudice.

The series of acoustic peaks in the CMB angular power spectrum is
strong evidence for the existence of primordial density perturbations
on super-horizon scales before the last-scattering of the CMB
photons.
The WMAP satellite has given further support by detecting an
anticorrelation on degree scales in the temperature-polarisation (TE)
cross-correlation, as expected for a spectrum of density perturbations
on super-horizon scales at last-scattering \citep{WMAP}.
The COBE satellite gave us the first direct measurement of the
amplitude of density fluctuations at last-scattering, and WMAP has now
taken another leap forward in determining not only the overall
amplitude but also the scale-dependence of the spectrum.

In this talk I will review how observations of primordial perturbation
spectra can constrain the parameters of different theoretical models
of the early universe.

\section{Origin of perturbations}

In the standard (radiation-dominated) hot big bang there is no way
to explain the origin of matter perturbations on scales greater
than the causal particle horizon, which is
simply related to the Hubble length at that time, $\ell_{\rm hor}
= cH^{-1}$.
A period of inflation 
changes the causal structure of the very
early universe. During inflation the causally connected region can
grow arbitrarily large.

Zero-point vacuum fluctuations of a free field $\phi$, with
comoving wavenumber $k\gg aH$, behave like an under-damped
harmonic oscillator:
 \be
 \delta\phi_k = \frac{1}{a\sqrt{2k}} e^{-ikt/a} \,,
 \ee
 where the overall normalisation is set by the canonical commutation
 relations. Linear evolution for a massless field during de Sitter
 inflation stretches these small-scale fluctuations up to large
 scales, $k\ll aH$, leading to a scale-invariant spectrum of
 perturbations on large scales
 \be
  \label{H2pi}
 \langle \delta\phi^2 \rangle = \left[ \frac{4\pi k^3}{(2\pi)^3}
 |\delta\phi_k^2| \right]_{k=aH} = \left( \frac{H}{2\pi} \right)^2
 \,.
\ee
More generally, a quasi-de Sitter expansion produces an almost
scale-invariant spectrum of fluctuations in any light, minimally
coupled scalar field.  Light fields, with an effective mass less than
the Hubble scale, become over-damped in the long-wavelength limit and,
as the decaying mode rapidly decays, the perturbations are well
described by a classical Gaussian random field. Equation~(\ref{H2pi})
then gives a estimate of the perturbation in the amplitude of the
asymptotic solution (as $k/aH\to0$) in terms of quantities at
Hubble-crossing ($k=aH$). Heavy fields, with mass greater than the
Hubble scale, remain under-damped with a steep blue ($k^3$) spectrum,
producing no classical perturbations on large scales.

%
Zero-point quantum fluctuations in the free gravitational field,
stretched by inflation to super-Hubble scales, yield a nearly
scale-invariant spectrum of gravitational waves. These tensor
modes remain decoupled from matter perturbations (to first-order) and
thus provide a direct probe of the inflationary dynamics. On large
scales (super-Hubble during the radiation era) we have
 \be
\label{T}
\langle T^2 \rangle
 = \frac{16}{\pi} \left( \frac{H}{M_{\rm Pl}} \right)^2
 \,.
 \ee
If we can detect tensor perturbations in the CMB anisotropies then
this would be a direct probe of the expansion rate during
inflation, but as yet there is no evidence that gravitational waves
contribute to the observed anisotropies of the CMB.

A detection of a spectrum of gravitational waves would be a great
triumph for inflation - a genuine prediction before the fact - but
there is no guarantee that gravitational waves will be produced at a
detectable level.
The amplitude of gravitational waves is a direct measure of how close
inflation occurs to the Planck scale, $M_{\rm Pl}$.
In models driven by a slowly rolling scalar field the
expansion rate is principally determined by the potential energy
of the field
 \be
\langle T^2 \rangle \simeq
 \frac{128}{3}\left( \frac{V}{M_{\rm Pl}^4} \right)
 \,.
 \ee
 Such a spectrum of primordial gravitational waves could be
 distinguished from scalar-type perturbations via B-mode polarisation
 of the CMB. For example, \citet{Lloyd} have argued that
 intrinsic B-mode polarisation from gravitational waves at
 last-scattering would be distinguishable from that generated by
 secondary effects only $V^{1/4}>3.2\times10^{15}$GeV. This is close
 to GUT scales and certainly possible in some models of inflation,
 but inflation could take place at much lower energy scales.


\section{Primordial density perturbation from inflation}

Around the time of last-scattering the cosmic fluid is composed of (at
least) photons, baryons, neutrinos and cold dark matter.
The primordial density perturbation can be characterised by an
overall density/curvature perturbation
%
 \be
\label{R}
 R = H \frac{\delta\rho}{\dot\rho}
 \,,
 \ee
and a relative density/isocurvature perturbation in the different
matter components
 \be
 S_m = 3H \left( \frac{\delta\rho_\gamma}{\dot\rho} -
 \frac{\delta\rho_m}{\dot\rho_m} \right) \,,
 \ee
 which describes the perturbed matter-to-photon number ratio,
 $n_m/n_\gamma$. Thus $S_m$ vanishes for the commonly considered case
 of a purely adiabatic primordial density perturbation. But in
 general it is possible to have isocurvature perturbations and these
 may be correlated with the curvature perturbations. We define the
 correlation angle
\be
\label{defTheta}
 \cos\Theta \equiv \frac{\langle RS \rangle}{\langle R^2\rangle^{1/2}
  \langle S^2\rangle^{1/2}} \,.
\ee

If there is only one light, slowly-rolling scalar field during
inflation (the {\em inflaton}) then Eq.~(\ref{H2pi}) describes an overall
density/curvature perturbation
\be
 \langle R^2 \rangle = \langle \left( \frac{H\delta\phi}{\dot\phi}
 \right)^2 \rangle_{k=aH}
\ee
which remains constant for adiabatic perturbations on large scales.
%
The existence of an almost scale-invariant spectrum of adiabatic
density perturbations is a generic prediction of single-field
inflation models. The amplitude of the perturbations is not in general
predicted, but rather matching the observed amplitude of density
perturbations on large scales is imposed as a constraint on the model
parameters.

In general, multiple scalar fields, $\varphi_i$, can produce both
curvature and isocurvature perturbations. During inflation,
analogously to the primordial era, we define instantaneous adiabatic
and entropy field perturbations \citep{GWBM}
\be 
 R_* = H \frac{\delta\phi}{\dot\phi}
\,,\qquad 
 S_* = H \frac{\delta\chi}{\dot\phi}
 \,,
\ee
where $\phi$ describes the evolution along the background
(homogeneous) trajectory in field space, and $\chi$ is orthogonal to
it. Of course one can have as many isocurvature perturbation modes
during inflation as one has additional light fields, but for
simplicity I will consider only two fields.
I have picked the (arbitrary) normalisation of the isocurvature
perturbation during inflation so that $R_*$ and $S_*$ have the same power
at Hubble-crossing, $\langle R_*^2 \rangle = \langle S_*^2 \rangle$,
although they are independent random fields $\langle R_*S_* \rangle=0$.

We can descibe the evolution of curvature and isocurvature
perturbations from Hubble-crossing during inflation to primordial
density perturbation via a transfer matrix \citep{Amendola,WBMR}
 \be
 \left( \begin{array}{c}R\\S\end{array} \right)
 = \left( \begin{array}{cc}1&T_{RS}\\0&T_{SS}\end{array} \right)
 \left( \begin{array}{c}R_*\\S_*\end{array} \right)
 \ee
where $T_{RS}$ and $T_{SS}$ are functions of $k$ to be
determined from the (observable) primordial perturbations:
\bea
T_{SS} &=& {\langle S^2 \rangle} / {\langle R^2 \rangle} \,,\\
T_{RS} &=& \cot\Theta \,,
\eea
where $\Theta$ is defined in Eq.(\ref{defTheta}).
$T_{SS}$ determines the amplitude of the surviving isocurvature
perturbation in the primordial era, and
$T_{RS}$ quantifies how much of the primordial curvature perturbation
is due to non-adiabatic field perturbations during inflation. 



\subsection{Scale-dependence}

The weak time-dependence (relative to the Hubble rate) of quantities
calculated at Hubble-crossing, $k=aH$, during slow-roll inflation yields a
weak scale-dependence in the spectra after inflation: 
\be
 \Delta n_x \equiv \frac{d\ln \langle x^2 \rangle}{d\ln k}
 \simeq \left( H^{-1}
  \frac{d\ln \langle x^2 \rangle}{dt} \right)_{k=aH} \,,
\ee
where the right-hand-side is evaluated at Hubble crossing.

Taking the logarithmic derivative of tensor power spectrum (\ref{T})
we obtain 
\be
 \Delta n_T
 \simeq -2\epsilon \,.
\ee
This is always negative and determined by the slow-roll parameter
\be
\epsilon \equiv -\frac{\dot{H}}{H^2} \simeq \frac{M_{\rm Pl}^2}{16\pi}
\left( \frac{V'}{V} \right)^2 \,.
\ee

This same slow-roll parameter appears as the ratio between the
amplitude of tensor perturbations (\ref{T}) to scalar curvature
perturbations (\ref{R}) at Hubble-crossing:
\be
\frac{\langle T^2 \rangle}{\langle R_*^2 \rangle} 
 = 16\epsilon \,.
\ee
This holds regardless of the number of light fields present during
 inflation. 
In terms of the observable (primordial) tensor and scalar perturbations
and the tensor tilt we have a two-field consistency relation
\citep{WBMR} 
\be
 \frac{\langle T^2 \rangle}{\langle R^2 \rangle} = - 8\Delta n_T
 \sin^2\Theta \,.
\ee

Taking the derivative of the three scalar perturbation spectra
(curvature, isocurvature and their cross-correlation) we obtain \citep{WBMR}
\footnote
{Note that the scale-dependence of the primordial scalar curvature
perturbation is more usually written as $n=1+\Delta n_R$.}
\bea
 \Delta n_R &\simeq& -(6-4\cos^2\Theta)\epsilon \nonumber\\
&&\ +
 2( \eta_{\phi\phi}\sin^2\Theta + 2 \eta_{\phi\chi}\sin\Theta\cos\Theta
 + \eta_{\chi\chi}\cos^2\Theta ) \,,\\
 \Delta n_S &\simeq& -2\epsilon + 2 \eta_{\chi\chi} \,,\\
 \Delta n_{RS} &\simeq& -2\epsilon + 2 \eta_{\chi\chi} +
 2\eta_{\phi\chi}\tan\Theta \,,
\eea
where the dimensionless mass matrix for the fields is given by
three more slow-roll parameters
\be
 \eta_{ij} = \frac{M_{\rm Pl}^2}{8\pi V} \frac{\partial^2
   V}{\partial\varphi_i\partial\varphi_j} \,.
\ee
Although the overall amplitude of $T_{RS}$ and $T_{SS}$ depends upon the
details of reheating and other physical processes from the end of
inflation up until the primordial (radiation-dominated) era, their
the scale dependence is determined by the evolution around
Hubble-crossing during inflation \citep{WBMR}.

\subsection{Inflaton scenario}

If the primordial curvature perturbations results solely from the
overall density/curvature perturbation at Hubble-crossing during
inflation (i.e., $T_{RS}\ll1$) then we have 
\be
\langle R^2\rangle = \langle R_*^2\rangle \,,
\ee
and hence we have the usual single-field tensor-scalar consistency
relation 
\be
 \frac{\langle T^2 \rangle}{\langle R^2 \rangle} = 16 \epsilon = -
 8\Delta n_T  \,.
\ee

The tilt of the primordial curvature perturbation is given by
\be
  \Delta n_R \simeq -6\epsilon + 2\eta_{\phi\phi} \,,
\ee

Any isocurvature perturbation must arise from a second light
field during inflation (e.g., the axion) and is uncorrelated with the
curvature perturbation ($\cos\Theta=0$). The spectral tilt of such an
isocurvature spectrum is given by
\be
 \Delta n_S \simeq -2\epsilon + 2 \eta_{\chi\chi} \,.
\ee
Because the curvature and isocurvature perturbations originate from
completely unrelated fields there is, a priori, no reason to expect
them to have comparable amplitude, i.e., $T_{SS}\sim1$, which is
required for primordial isocurvature modes to be detectable.

\subsection{Curvaton scenario}

In the curvaton scenario \citep{ES,LW,MT} the primordial density perturbation
is supposed to arise from quantum fluctuations during inflation in a
light scalar field, other than the inflaton, which then decays some
time after inflation. It is one example of a model where the primordial
perturbation comes entirely from an isocurvature
perturbation at Hubble-crossing (i.e, $T_{RS}\gg1$):
\be 
\langle R^2\rangle = T_{RS}^2 \langle S_*^2\rangle \,.  
\ee 
The curvature perturbation at Hubble-exit during inflation is then
negligible, and the tensor-scalar consistency relation becomes
\be
 \frac{\langle T^2 \rangle}{\langle R^2 \rangle} \ll - 8\Delta n_T
 = 16 \epsilon < 16\,.
\ee
The curvaton scenario provides a class of inflationary models
where gravitational waves must have a negligible effect on the CMB
anisotropies. 

The scalar tilt in the curvaton scenario is given by
\be
 \Delta n_R \simeq -2\epsilon + 2\eta_{\chi\chi} \,,
\ee
Because the original perturbations that seed the primordial curvature
perturbation are non-adiabatic at Hubble-exit during inflation it is
possible for the curvaton to leave {\em residual} isocurvature perturbations
after the curvaton decays, depending upon the relative timescales of
the matter-decoupling and curvaton decay \citep{LUW}. In any case the
curvature and residual isocurvature perturbations must be completely
correlated ($\cos\Theta=1$) and share the same spectral tilt
$ \Delta n_R = \Delta n_S$.
%
There is no lower limit on the amplitude of the primordial
isocurvature perturbation (it could be negligible) but
the curvaton does provide a simple physical model where primordial
curvature and isocurvature perturbations have similar amplitude. In
the simplest such case, where the matter decouples while the curvaton
density is negligible, before the curvaton decays, then we find
$T_{SS}=3T_{RS}$, yielding $S=3R$ \citep{LUW}. Unfortunately such a large
isocurvature perturbation is no longer compatible with current CMB
data when considering cold dark matter or baryon isocurvature
modes \citep{GL}.

\subsection{Scale-dependent tilt?}

If the WMAP data as yet gives no hint of gravitational waves or
non-adiabatic modes, it offers the tantalising prospect of detecting
some scale-dependence of the scalar tilt, $d\Delta n_R/d\ln k$
\citep{WMAP}.  Combining CMB data on large scales with
survey data on smaller scales may favour a spectrum with less power
on very large scales (suggested by the ``low quadrupole'' of the CMB)
and less power on smaller scales than expected for a scale-invariant
spectrum. This may not fit that easily with our models of slowly
evolving, slow-roll inflation.

In the single-field scenario the running of the tilt is given
in the slow-roll approximation by
\be
\frac{d\Delta n_R}{d\ln k} \simeq - 2\xi_{\phi\phi}
-8\epsilon(3\epsilon-2\eta_{\phi\phi}) \,,
\ee
while in the curvaton scenario one obtains
\be
\frac{d\Delta n_R}{d\ln k} \simeq - 2\xi_{\chi\phi}
-4\epsilon(2\epsilon-\eta_{\phi\phi}-\eta_{\chi\chi}) \,,
\ee
The parameters
\bea
\xi_{\phi\phi} &=& \frac{M_{\rm Pl}^4}{64\pi^2V^2} \frac{\partial
  V}{\partial\phi} 
\frac{\partial^3V}{\partial\phi^3} \,, \\
\xi_{\chi\phi} &=& \frac{M_{\rm Pl}^4}{64\pi^2V^2} \frac{\partial
  V}{\partial\phi} \frac{\partial^3V}{\partial\phi \partial\chi^2} \,,
\eea
are a dimensionless measure of how fast the scalar field's effective mass
is changing on a Hubble time. In particular for the inflaton we have
\be
\frac{d\eta_{\phi\phi}}{dt} \simeq 2H \left( \epsilon\eta_{\phi\phi} -
  \xi_{\phi\phi} \right) \,.
\ee

In either inflaton or curvaton scenarios the $\xi$-parameters involve four
derivatives of the potential and are expected to be of the same order
as $\epsilon^2$. Hence the small tilt observed suggests that the running
should also be small. If not there is a new coincidence problem as to
why the primordial power spectrum should reach a maximum on scales
around $100$Mpc. 

In the inflaton case a value for $\xi_{\phi\phi}\sim\eta_{\phi\phi}$
suggests that slow-roll inflation may break down in a few Hubble
times. We may have to move away from the notion that inflation is a
slow-rolling, weakly time-dependent process, and
consider more transient bursts of inflation and/or generation of
perturbations. 
In the curvaton case $\xi_{\chi\phi}\sim\eta_{\chi\chi}$ would just
signal that the curvaton field becomes heavy in a few Hubble times,
which is not inconsistent with slow-roll inflation continuing.

\section{Discussion}

Increasingly precise constraints upon the spectral tilt and possible
running of the spectral tilt will place increasingly tight constraints
upon models of inflation. These are usually characterised by the
slow-roll parameters $\epsilon$, $\eta$ and $\xi$ which give a useful
qualitative understanding of inflationary dynamics. But an unexpectedly
large running of the tilt might be an indication of the limitations of
the slow-roll description.
Observations may require more sophisticated models of inflation if
their parameters can be related to observables. For instance
non-adiabatic perturbations produced when there is more than one light
field during inflation, provide an additional source of primordial
density perturbations. These might provide features in the primordial
power spectrum or non-Gaussianity without violating slow-roll.

Precise observations offer the hope of seeing more than just the
primordial curvature perturbation. Evidence of primordial
gravitational waves would be an important test of inflation. Even in
multi-field models of inflation there are consistency conditions
constraining the tensor-scalar ratio. Indeed a {\em large} amplitude
of gravitational waves could {\em falsify} any slow-roll model of
inflation.
Primordial isocurvature modes and/or non-Gaussianity would also
provide valuable additional information about the physics of
inflation and the very early universe.


{\em Acknowledgements}\\
I'm grateful to Nicola Bartolo, David Lyth, Karim Malik, Sabino
Matarrese, Toni Riotto and Carlo Ungarelli for the collaborations upon
which this talk is based.
My work is supported by a Royal Society University Research
Fellowship.




\end{document}